\documentclass{article}
\pdfoutput=1

\PassOptionsToPackage{numbers}{natbib}

\usepackage[final]{neurips_2024}

\usepackage[utf8]{inputenc} 
\usepackage[T1]{fontenc}    
\usepackage{hyperref}       
\usepackage{url}            
\usepackage{booktabs}       
\usepackage{amsfonts}       
\usepackage{xfrac}       
\usepackage{microtype}      
\usepackage{xcolor}         

\usepackage{graphicx}
\usepackage{array}
\usepackage{tabularx}
\usepackage{geometry}
\usepackage{longtable}
\usepackage{float}
\usepackage{enumitem}
\usepackage{placeins}
\usepackage{multicol}
\usepackage{amsmath}
\renewcommand{\arraystretch}{1.1}

\setlist[itemize]{left=0pt, itemsep=0pt, topsep=0pt, parsep=0pt, partopsep=0pt}

\title{What you say or how you say it? Predicting Conflict Outcomes in Real and LLM-Generated Conversations}

\author{
    Priya Ronald D'Costa \\
  University of Pennsylvania\\
  \texttt{pdcosta@seas.upenn.edu} \\
  \and
  \textbf{Evan Rowbotham} \\
  New York University \\
  \texttt{err6934@nyu.edu} \\
  \and
  \textbf{Xinlan Emily Hu} \\
    University of Pennsylvania\\
  \texttt{xehu@wharton.upenn.edu}
}

\begin{document}
\maketitle

\begin{abstract}
When conflicts escalate, is it due to what is said or how it is said? In the conflict literature, two theoretical approaches take opposing views: one focuses on the content of the disagreement, while the other focuses on how it is expressed. This paper aims to integrate these two perspectives through a computational analysis of 191 communication features — 128 related to expression and 63 to content. We analyze 1,200 GPT-4 simulated conversations and 12,630 real-world discussions from Reddit. We find that expression features more reliably predict destructive conflict outcomes across both settings, although the most important features differ. In the Reddit data, conversational dynamics such as turn-taking and conversational equality are highly predictive, but they are not predictive in simulated conversations. These results may suggest a possible limitation in simulating social interactions with language models, and we discuss the implications for our findings on building social computing systems.
\end{abstract}

\section{Introduction}

Conflict is a double-edged sword: it can harm teams’ productivity and satisfaction, but when managed effectively, it can also enhance decision-making. \cite{Gladstein1984,Eisenhardt1990}. Scholars traditionally classify conflict into three types: task conflict, which relates to differences in opinions about tasks; relationship conflict, involving interpersonal disagreements; and process conflict, dealing with coordination issues \cite{Jehn1997,Behfar2011}. Task and process conflicts can sometimes be constructive, fostering diverse ideas and clarifying expectations, while relationship conflict is generally seen as detrimental \cite{Jehn2001}.

In practice, these conflict types often overlap. Conversations can easily shift between task-related and interpersonal disagreements, complicating the distinction between constructive and destructive conflicts. Interpersonal disagreements can “bleed” into disagreements over taskwork, making it difficult to disentangle the characteristics of “constructive” and “destructive” conflicts As a result, research has yielded mixed findings on the effects of different conflict types \cite{DeChurch2013,DeDreu2003,deWit2012,O'Neill2013}.

As an alternative to categorizing the content of disagreement, Weingart et al.~\cite{Weingart2015} propose focusing on how conflict is expressed. Specifically, they identify two attributes of conflict expression: (1) Directness and (2) Oppositional Intensity. Directness is defined as “the degree to which the sender explicitly versus implicitly conveys his or her opposition,” and Oppositional Intensity as “the degree of strength, force, or energy with which the sender conveys opposition during a given conflict event.” They propose that, depending on how conflict is expressed, the interaction can either escalate or de-escalate. For example, a very direct and oppositionally intense conflict statement (“I hate your idea and think you are stupid”) might lead to a more negative emotional response than a less direct or intense statement (“I am not sure I agree and wonder if we can reconsider”).

Critically, however, the two theoretical approaches to conflict have not been reconciled: is the primary driver of conflict what people say, or how people say it? In this paper, we present a computational analysis of the expressions and content of conflict exchanges. Across two datasets — a set of conversations generated by GPT-4, and a set of conversations on Reddit — we examine which language markers are most associated with “constructive” versus “destructive” outcomes. We then compare whether these markers of conflict expression are more predictive of conflict outcomes than features associated with the topic of conversation. Our analysis integrates the two theoretical approaches to conflict, and to provide an initial answer to how the content and expression of a conflict each contribute to how conflict is resolved.

Our approach builds on significant previous work in both computer and social science, which seeks to quantify attributes of conversations that predict meaningful collective outcomes. For example, Yeomans et al.~\cite{Yeomans2020} use 39 linguistic markers, such as acknowledgement, reassurance and apologies, to measure ``receptiveness,'' or the willingness to thoughtfully engage with opposing views. Similarly, Danescu-Niculescu-Mizil et al.~\cite{Danescu-Niculescu-Mizil2013} in the ConvoKit package identify politeness in conversations; Islam et al.~\cite{Islam2020} examine a speaker’s lack of commitment (hedging); Gray et al.~\cite{Gray2019} examine the extent to which successive messages in a conversation introduces new ideas; and Lix et al.~\cite{Lix2020} study the “discursive diversity” in subject matter contributed by different team members. In each of these cases, the authors demonstrate that quantifying language allows for the prediction of different psychologically-relevant outcomes, whether it is preventing a conflict from escalating (receptiveness; politeness), being able to generate more creative ideas (forward flow), or being more likely to deliver a software project on time (discursive diversity).

By using language to predict meaningful outcomes, our work further contributes to the social computing literature; for example, our comparative analysis of synthetic and “real” conversations holds implications for the types of conflict interactions that can be effectively simulated by language models (see related work in \cite{park2022socialsimulacracreatingpopulated,shaikh2024rehearsalsimulatingconflictteach,liu2023improvinginterpersonalcommunicationsimulating}). Additionally, our work connects with research on the live detection and intervention of conflict on social computing systems (for example, \cite{Chang_2022}). Taken together, this research both advances the theoretical conversation on modeling conflict, as well as generates practical applications for social computing and human-computer interaction.

\section{Methods}

\subsection{Data}

\textbf{Data Sources:} We compare two sources of data across our experiments. The first is a collection of 1,200 synthetic conflict conversations generated by GPT-4 (“Synthetic”), and the second is a collection of 12,630 online discussions from the “Change My View” Subreddit (“Reddit”), which were previously collected in \cite{Tan2016} and \cite{Chang2019}. These two data sources serve different purposes; the former ensures construct validity by generating “controlled” instances of constructive and destructive conflicts, all else equal; the second ensures ecological validity by capturing real conversations on meaningful topics. Refer to Appendix~\ref{gpt-prompts} for details on the prompts used to generate the synthetic datasets and to Appendix~\ref{reddit-data-processing} for details on pre-processing the online discussions. 

\textbf{Dependent Variable:} We define the outcome of a conversation as either “destructive” or “constructive” in each dataset as follows. In Synthetic conversations, we explicitly prompt GPT to generate a particular outcome (see Appendix~\ref{gpt-prompts}); in the Reddit data, we use a subset of conversations that have devolved into personal attacks \cite{Chang2019}, which we consider to be “destructive,” and conversations that were awarded a “delta” — an indication that a person had changed their mind, which we considered to be the marker of a “constructive” conversation. Table~\ref{sample-conversation-table} (in Appendix~\ref{sample-data}) presents excerpts of destructive and constructive conversation from each of the two datasets.

\subsection{Models}

\subsubsection{Features}

We identify two general classes of conversational features: expression features capture aspects of “how” a person communicated, while content features capture aspects of “what” a person communicated. Features computed at the level of a single utterance are averaged across a conversation. Finally, we z-score all features, ensuring they have zero mean and unit standard deviation.

\textbf{Expression Features:} Using the Team Communication Toolkit \cite{Hu2024} (Appendix~\ref{tct}), a collection of previously-validated instruments for quantifying text, we generate 123 utterance-level attributes (measuring how each utterance is expressed, such as whether the speaker used positive sentiment or showed hesitation) and 5 conversation-level attributes (measuring dynamics of the overall interaction, such as whether airtime is shared equally and whether individuals take turns speaking).

\textbf{Content Features:} The Team Communication Toolkit also generates 32 semantic features (27 at the utterance level, and 5 at the conversation level), including both bag-of-words measures (e.g., LIWC’s “religion,” “work,” and “money” lexicons; \cite{pennebaker}) and vector-based measures of similarity (e.g., Discursive Diversity; \cite{Lix2020}).

We also curate a set of dummy-encoded topics specifically for the Reddit data. Using BERTopic~\cite{Grootendorst2022} we cluster the Reddit data into 30 conversation topics, which cover 65\% of the Reddit dataset. Unclassified conversations are assigned to a “residual topic” (see Appendix~\ref{topic-modeling} for details). A human researcher then summarized each topic into an interpretable label based on its representative words (for example, 'israel', 'palestinians', 'jews', 'palestine' can be summarized as “Israel-Palestine”).

\textbf{Validation of Conflict Expression Measurement:} We validated whether the extracted features capture theoretically relevant conflict expressions (Directness and Oppositional Intensity). For Synthetic data validation, we asked GPT-4 to generate 1,200 balanced utterances that are either high or low in Directness, and either high or low in Oppositional Intensity. Using the 123 utterance-level features from the Team Communication Toolkit, our XGBoost model achieves an F1 score of 0.93 for Directness and 0.94 Oppositional Intensity, which we take to be strong evidence that our features capture relevant indicators of conflict expression. To validate conflict expressions in the Reddit data, we asked three trained research assistants to annotate 929 utterances from 121 conversations randomly sampled from the Reddit data. Each utterance was rated as either high or low in Directness, and high or low in Oppositional Intensity (for more details on annotation, refer to Appendix~\ref{human-annotation-proc}). We resample the dataset to ensure class balance by selecting 500 samples (with replacement; 250 labeled as 'high' and 250 labeled as 'low' for Directness of expression). We apply the same resampling process to obtain a balanced set for Oppositional Intensity of expression, and use these balanced datasets to replicate our findings on the Synthetic data: our XGBoost model achieved an F1 score of 0.95 for Directness and 0.94 for Oppositional Intensity.

\subsubsection{Modeling Approach}

Across all experiments, we fit vanilla Extra Gradient Boosted Trees (XGBoost classifiers, \cite{Chen2016}) to predict the dependent outcome (whether the conversation was “constructive” or “destructive”). We chose XGBoost in order to account for possible nonlinear interactions among conversational attributes, but our general approach can be replicated with other modeling schemes or improved with hyperparameter optimization.

Specifically, we evaluate six models that probe different combinations of expression- or content-based features: (1) Expression Features Only; (2) Content Features – Semantic Only (the 32 features generated by the Team Communication Toolkit); (3) Content Features – Topic Only (the 31 topic labels generated by BERTopic); (4) All Content Features; (5) All Team Communication Toolkit Features (combining Expression and Content-Semantic); and (6) All Features.

\subsubsection{Evaluation}

We use the F1 score to measure model performance on a held out 20 of the data. This provides a signal of whether expression- or content-based features more effectively predict unseen conflict outcomes when considered in isolation (Models 1 - 4). Second, we use SHAPley values \cite{Lundberg2017} to examine feature importance, enabling us to determine whether expression or content features are more important when they are included as part of the same model (Models 5 - 6). See Appendix~\ref{main_results} for additional details.

\section{Results}

We present our results in Appendix~\ref{main_results}, with primary metrics in Table~\ref{tab:main-results} and feature importance details in Appendices~\ref{feature-importance-synthetic} (for the Synthetic data) and~\ref{feature-importance-reddit} (for the Reddit data). We present the F1 scores and top five features for each of our models across both datasets. We note that, since we did not generate topic features for the Synthetic dataset, Models 3, 4, and 6 apply only to the Reddit data.

Across both datasets, F1 on the held-out data was higher for Model 1 (Expression Only; $F1_{Synthetic}$ = 0.98, $F1_{Reddit}$ = 0.88) than for Models 2 - 4 (the three content feature-based models; in Model 2, $F1_{Synthetic}$ = 0.88, $F1_{Reddit}$ = 0.81; in Model 3,  $F1_{Reddit}$ = 0.59, in Model 4, $F1_{Reddit}$ = 0.81). 

Moreover, the Expression Only model performed at parity with models combining expression and content features (Models 5-6), suggesting that expression features are more predictive of conflict outcomes, and semantic features do not add predictive power ($F1_{Synthetic}$ = 0.98, $F1_{Reddit}$ = 0.88). 

Inspection of feature importance yields several interesting patterns. In Model 1, we observe patterns that align with our intuitions — for example, conversations that are more positive on average tend to be more constructive, and those that are more negative on average tend to be more destructive. Additionally, expressions of gratitude and acknowledging the other’s point of view tend to be positively associated with constructive conflicts.

However, in the Reddit data, the top predictors also include features associated with the overall conversation dynamics, such as the extent to which participants take turns rather than making many successive utterances (Turn-Taking Index \cite{Almaatouq2024}), the extent to which airtime in the conversation is shared equally (Gini Coefficient; \cite{Tausczik2013}), and the timing of messages. These features are not significant for the Synthetic data. GPT-4 tended to produce a regular alternating structure (in which one speaker makes a remark, and another responds in turn); on Reddit, however, speakers behaved irregularly (some users were substantially more outspoken than others). Moreover, in the Synthetic data, timestamp metadata was assigned arbitrarily, and thus contained no value. Features associated with equality, turn-taking, and timing therefore carried a more meaningful signal in the real Reddit conversations, while these features were meaningless in the Synthetic data. In this way, our analysis reveals richness in real-world conversations that are not reflected in a simulated conversation.

The topics alone were the least predictive of conflict outcomes, with only one topic (Race and Donald Trump) indicating a strong likelihood of a destructive conflict. Among the other measures of content, we find that discursive diversity (a measure of semantic divergence among speakers) is associated with more destructive conflicts, particularly on Reddit. This suggests that, when speakers are not discussing the same ideas (perhaps talking past each other), the conflict is more likely to end poorly.

\section{Discussion and Conclusion}

In this paper, we integrate two theoretical approaches to conflict — one focusing on the content (what is said) and the other focusing on the expression (how it is said). By quantifying content- and expression-based features across two datasets, we contribute to the methodology of studying conflict, answering Bendersky et al.’s call~\cite{Bendersky} to better specify and measure the “features of conflict episodes.” Furthermore, by using these features to predict conflict outcomes, we investigate how content and expression contribute differently to dynamics of escalation and de-escalation.

Our results suggest that, at least within the contexts studied, expression-based attributes are more predictive of conflict outcomes, but the specific features depend heavily on the context. Whereas, in simulated conflicts, the most informative features were associated with emotions and politeness, in real Reddit conflicts, the informative features include turn-taking, equality, and other features capturing dynamic interactions between speakers. This finding may reveal a limitation in a language model’s ability to fully capture realistic conflict dynamics, and align with previous observations that generated conversations appear to be overly formal \cite{park2023generativeagentsinteractivesimulacra}. Consequently, while there has been considerable promise in using generative agents to simulate conflict dynamics (\cite{park2022socialsimulacracreatingpopulated}; \cite{shaikh2024rehearsalsimulatingconflictteach}), some aspects of conflict may not be adequately captured.

The finding that expression-based attributes are highly predictive of conflict outcomes is also promising, as it suggests that it is possible to identify destructive conflict while remaining agnostic to the content of the discussion. Thus, we see early detection and intervention mechanisms along the lines of \cite{Chang_2022} as fruitful avenues of exploration.

However, our results are limited to only two specific contexts: a set of conversations simulated by GPT, and a set of conversations on Reddit. Neither of these contexts fully captures the richness and variety of the environments in which conflicts take place. Social media, for example, affords very different interactions compared to the workplace or the home, and the particular subreddit we study, r/ChangeMyView, establishes specific communication norms that may not generalize to other environments. Thus, future work should seek to generalize our findings to other settings, performing a true “out of sample” (or even “out of distribution”) test. It should also explore different modalities beyond text — audio cues, such as one’s tone or volume, and visual cues, such as facial expressions and head nods, are likely valuable signals of conflict expression. Weingart et al.~\cite{Weingart2015} even propose that conflict is expressed through what is left unsaid: for example, choosing not to engage with a person may be a possible method of indirectly expressing conflict.

Future work should also probe how attributes of the environment interact with communication attributes. Here, we consider content and expression features largely as orthogonal aspects of communication. In reality, however, these attributes may interact; a particularly sensitive topic might prompt specific patterns of expression — such as expressing more hesitation — which in turn might lead to different patterns of conflict escalation and de-escalation.

Overall, however, our work represents a meaningful first step in exploring these relationships, shedding light on how “what you say” and “how you say it” can make or break a disagreement.

\bibliography{references}

\section{Appendices}
\subsection{Main Results}~\label{main_results}

The table presents the evaluation metric (F1) for each model on the held-out set, and is followed by the SHAPley value \cite{Lundberg2017} plots and their interpretation.

The models are as follows:
\begin{enumerate}
    \item \textbf{Expression Features Only}
    \item \textbf{Content Features – Semantic Only} (the 32 features generated by the Team Communication Toolkit)
    \item \textbf{Content Features – Topic Only} (the 31 topic labels generated by BERTopic)
    \item \textbf{All Content Features} (Semantic and Topic)
    \item \textbf{All Team Communication Toolkit Features} (Expression and Content - Semantic)
    \item \textbf{All Features}
\end{enumerate}

The first row indicates the number of features that were included in the model. The second row indicates whether or not the model was regularized (in cases with a large number of combined Content and Expression features, we used Powershap~\cite{Verhaeghe2022} at $\alpha = 0.01$ to regularize features prior to fitting the vanilla XGBoost model.

\begin{table}[ht]
\centering
\begingroup\small
\renewcommand{\arraystretch}{1.1} 

\begin{longtable}{>{\centering\arraybackslash}p{3cm}
                >{\centering\arraybackslash}p{1cm}
                >{\centering\arraybackslash}p{1cm}
                >{\centering\arraybackslash}p{1cm}
                >{\centering\arraybackslash}p{1cm}
                >{\centering\arraybackslash}p{1cm}
                >{\centering\arraybackslash}p{1cm}}

\hline
\textbf{} & \textbf{(1)} & \textbf{(2)} & \textbf{(3)} & \textbf{(4)} & \textbf{(5)} & \textbf{(6)} \\
\textbf{All Features} \\
\hline
\endhead

\hline
\endfoot

\hline
\endlastfoot

Original Number of Features & 128 & 32 & 31 & 63 & 160 & 191 \\
\hline
Regularized? & No & No & No & Yes & Yes & Yes \\
\hline
F1 Score (Synthetic) & 0.98 & 0.88 & N/A & N/A & 0.98 & N/A \\
\hline
F1 Score (Reddit) & 0.88 & 0.81 & 0.59 & 0.81 & 0.88 & 0.88 \\
\hline

\caption{The primary results across our five models. We present the F1 scores for each model on both the Reddit and Synthetic datasets, as well as the top five features by their SHAPley values. Each column shows the results of a different model specification: (1) Expression Features Only; (2) Content Features – Semantic Only (the 32 features generated by the Team Communication Toolkit); (3) Content Features – Topic Only (the 31 topic labels generated by BERTopic); (4) All Content Features; (5) All Team Communication Toolkit Features (combining expression and non-topic-related semantic features); and (6) All Features. We note that, since we did not generate topic features for the Synthetic dataset, Models 3, 4, and 6 apply only to the Reddit data.}
\label{tab:main-results}
\end{longtable}
\endgroup
\end{table}

\subsubsection{Feature Importance: Synthetic Dataset}\label{feature-importance-synthetic}
Here, we present detailed results of the feature importance of each model, showing the SHAPley value plot~\cite{Lundberg2017} for each model. 

SHAPley value plots can be read as follows: Each point on the plot represents a single conversation, and the color of the point represents whether the feature value was high (red) or low (blue). The position of the point on the x-axis represents the impact of the feature on the outcome; points far to the left can be interpreted as significant in predicting that a conversation was destructive, while points far to the right can be interpreted as significant in predicting that a conversation was constructive. Thus, a cluster of red points far to the right can be interpreted as showing that a high value for a given feature positively correlates with a constructive conflict. Finally, from top to bottom on the y-axis, feature names are ordered from most important to least important.

We specifically highlight the top five features for each model for the Synthetic data, and present them in a format as follows:

\begin{quote}
    Feature Name (direction of impact; +/-)
\end{quote}

\textbf{(+)} indicates that a high value of a feature \textit{increases} the chances that a conflict outcome is contstructive; \textbf{(-)} indicates that a high value of a feature \textit{decreases} the changes that a conflict outcome is constructive (in other words, it is more likely to be destructive).

\subsubsection*{(Model 1) Expression Only}

Top 5 features by SHAPley values:
\begin{itemize}
    \item RoBERTa Positivity (+)
    \item RoBERTa Negativity (-)
    \item Proportion of First Person Pronouns (+)
    \item Receptiveness Markers - Acknowledgement (+)
    \item Politeness Strategies - 1st Person Start (+)
\end{itemize}

In the Synthetic data, the top features are associated with the sentiment of a conversation; unsurprisingly, more positive conversations tend to be more constructive, while more negative conversations tend to be more destructive. The next three features are markers of politeness and receptiveness, such as the proportion of first-person pronouns and use of acknowledgement. All of these are positively associated with a constructive outcome. 

\begin{figure}[ht]
    \centering
    \includegraphics[width=0.9\textwidth,height=0.5\textheight]{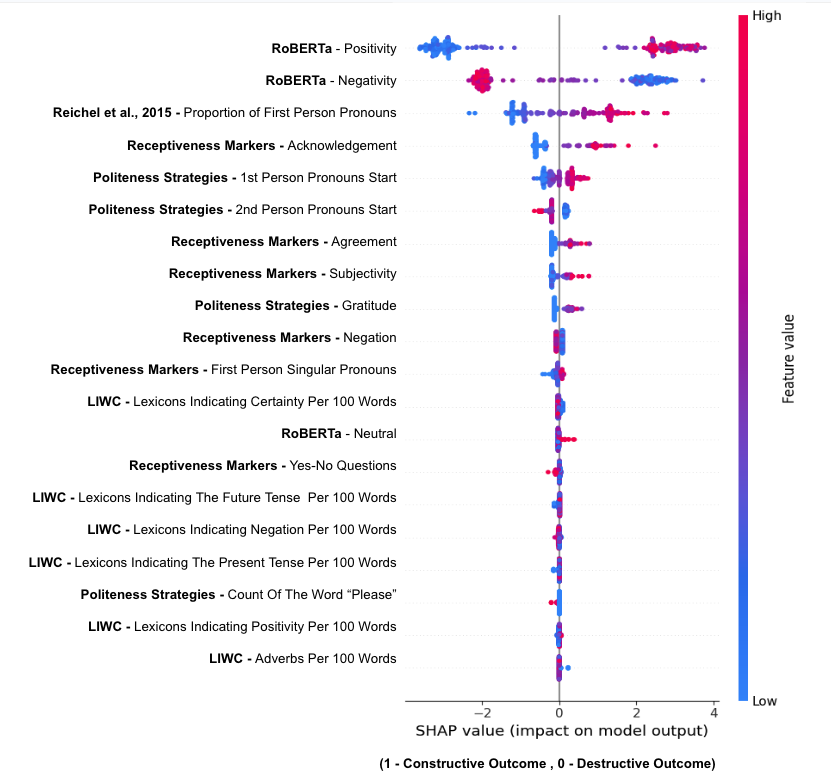} 
    \caption{Synthetic Data: Expression Only} 
    \label{fig:synthetic_1} 
\end{figure}

\clearpage 

\subsubsection*{(Model 2) Content (Semantic Only)}

Top 5 features by SHAPley values:
\begin{itemize}
    \item LIWC - Insights (+)
    \item Content Word Accommodation (-)
    \item LIWC - Relative (+)
    \item Information Exchange Z-Score (Chats) (-)
    \item LIWC - Anxiety (+)
\end{itemize}

Use of words from the ``Insight,'' ``Relative,'' and ``Anxiety'' lexicons are positively associated with a constructive outcome; speaking very similarly to the person in the previous turn (high Content Word Accommodation) and use of more ``information'' (content) words are associated with a more destructive outcome.

\begin{figure}[ht]
    \centering
    \includegraphics[width=0.9\textwidth,height=0.5\textheight]{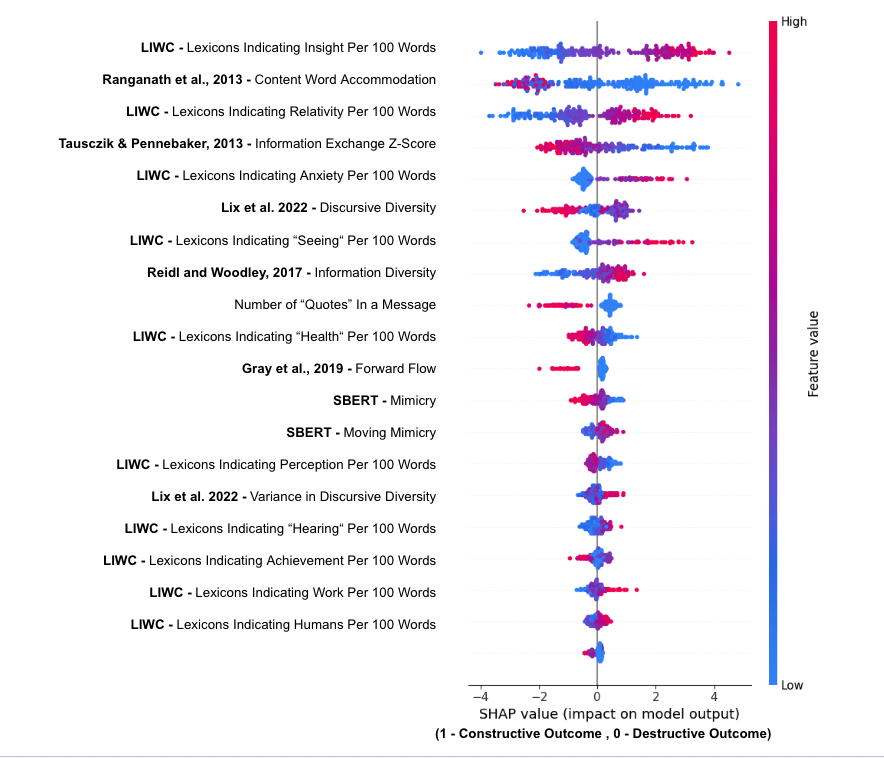} 
    \caption{Synthetic Data: Content (Semantic Only)} 
    \label{fig:synthetic_3} 
\end{figure}

\clearpage 

\subsubsection*{(Model 5) All TCT (Expression + Semantic)}

Top 5 features by SHAPley values:
\begin{itemize}
    \item RoBERTa Positivity (+)
    \item RoBERTa Negativity (-)
    \item Proportion of First Person Pronouns (+)
    \item Receptiveness Markers - Acknowledgement (+)
    \item Receptiveness Markers - Agreement (+)
\end{itemize}

Here, the results are very similar to those of Model 1.

\begin{figure}[ht]
    \centering
    \includegraphics[width=0.8\textwidth,height=0.3\textheight]{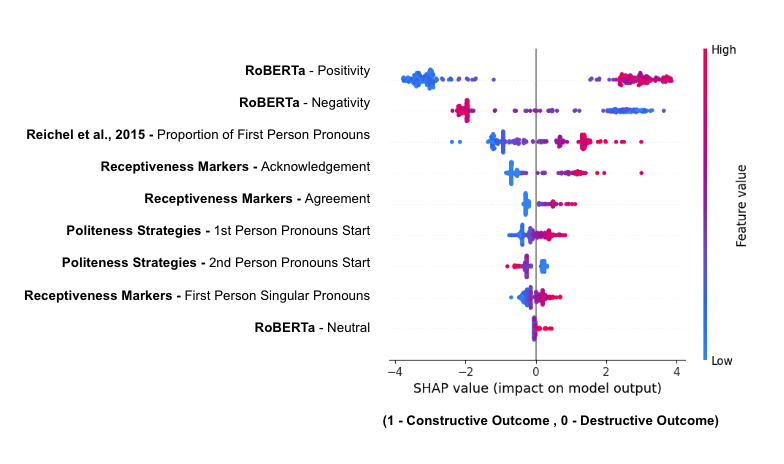} 
    \caption{Synthetic Data: All Team Communication Toolkit Features} 
    \label{fig:synthetic_5} 
\end{figure}

\clearpage 
\subsubsection{Feature Importance: Reddit Dataset}\label{feature-importance-reddit}

We specifically highlight the top five features for each model for the Reddit data, presenting them in the same manner as we did for the Synthetic data.

\subsubsection*{(Model 1) Expression Only}
Top 5 features by SHAPley values:
\begin{itemize}
    \item Turn Taking Index (-)
    \item Gini Coefficient - Sum of Number of Messages (-)
    \item RoBERTa - Negative (-)
    \item Time Difference (-)
    \item Politeness Strategies - Gratitude (+)
\end{itemize}

A high degree of turn-taking and a more unequal conversation (in which fewer individuals dominate the conversation) are associated with more destructive outcomes. Longer time differences between messages are also generally associated with more destructive outcomes. As before, negative sentiment is associated with more destructive outcomes, but markers of politeness are associated with more constructive outcomes.

\begin{figure}[ht]
    \centering
    \includegraphics[width=0.9\textwidth,height=0.5\textheight]{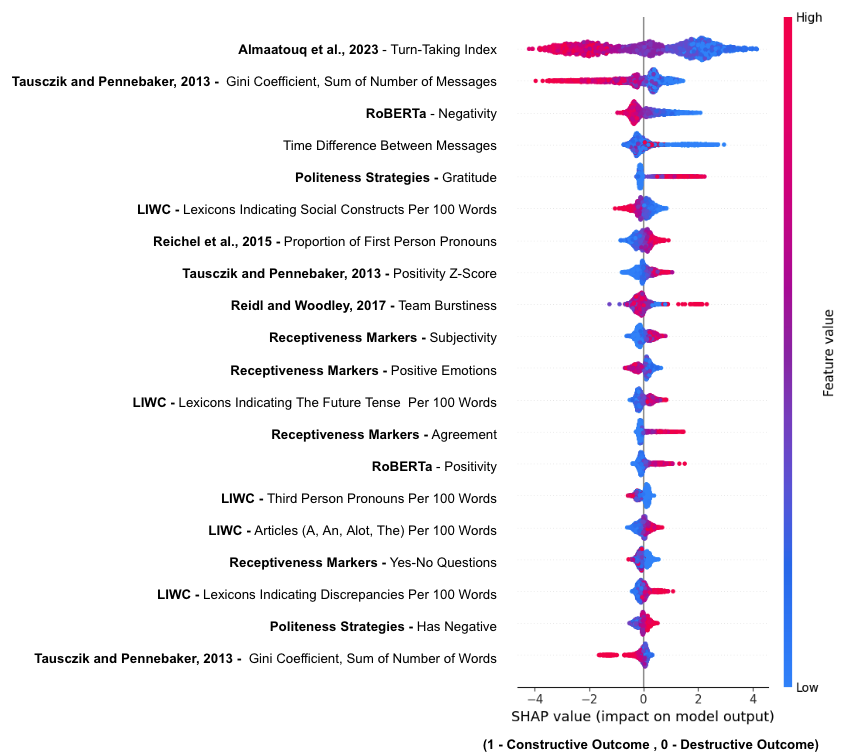} 
    \caption{Reddit Data: Expression Only} 
    \label{fig:real_1} 
\end{figure}

\clearpage 

\subsubsection*{(Model 2) Content (Semantic Only)}
Top 5 features by SHAPley values:
\begin{itemize}
    \item Variance in Discursive Diversity (-)
    \item Discursive Diversity (-)
    \item Mimicry - BERT (-)
    \item LIWC - Relative (+)
    \item Within Person Discursive Range (-) 
\end{itemize}

Here, a high variance in Discursive Diversity (the extent to which Discursive Diversity varies across the course of a conversation) is associated with more destructive conflicts, and high Discursive Diversity (the semantic distance between the contributions of each speaker in the conversation) and high Within-Person Discursive Range (variance in semantic meaning by the same person over the course of a conversation) are associated with more destructive conversations. High semantic similarity between successive messages is also associated with more destructive conflicts. Words from the ``Relative'' lexicon are associated with more constructive conflicts.

\begin{figure}[ht]
    \centering
    \includegraphics[width=0.9\textwidth,height=0.5\textheight]{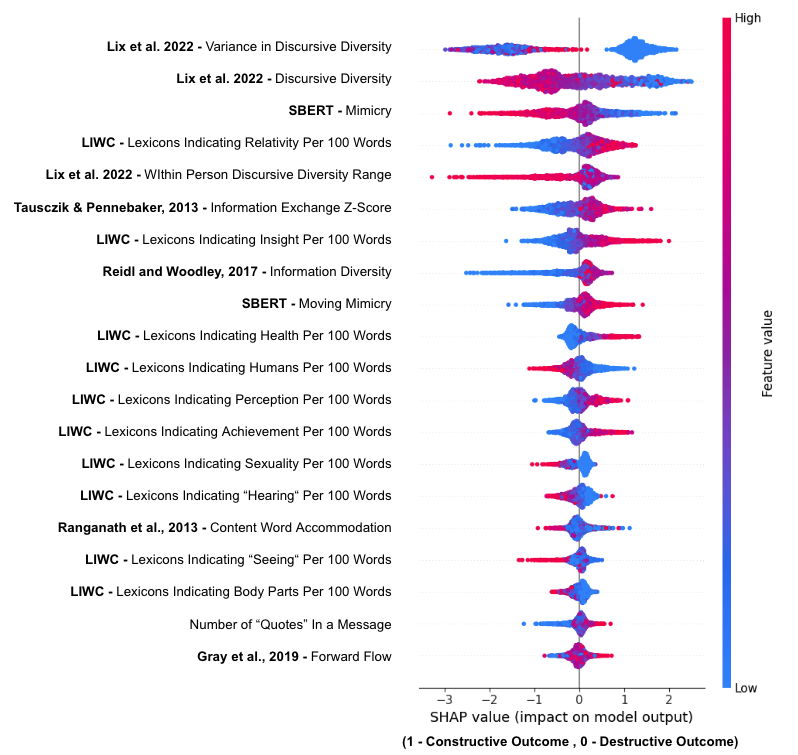} 
    \caption{Reddit Data: Content (Semantic Only)} 
    \label{fig:real_2} 
\end{figure}
\clearpage 

\subsubsection*{(Model 3) Content (Topic Only)}
Top 5 features by SHAPley values:
\begin{itemize}
    \item Race and Donald Trump (-)
    \item Residual Topic (\textit{n.s.})
    \item Gender (\textit{n.s.})
    \item Education (\textit{n.s.})
    \item Drugs and Alcohol (\textit{n.s.})
\end{itemize}

The only significant topic feature is ``Race and Donald Trump,'' which was associated with more destructive conflicts.

\begin{figure}[ht]
    \centering
    \includegraphics[width=0.9\textwidth,height=0.5\textheight]{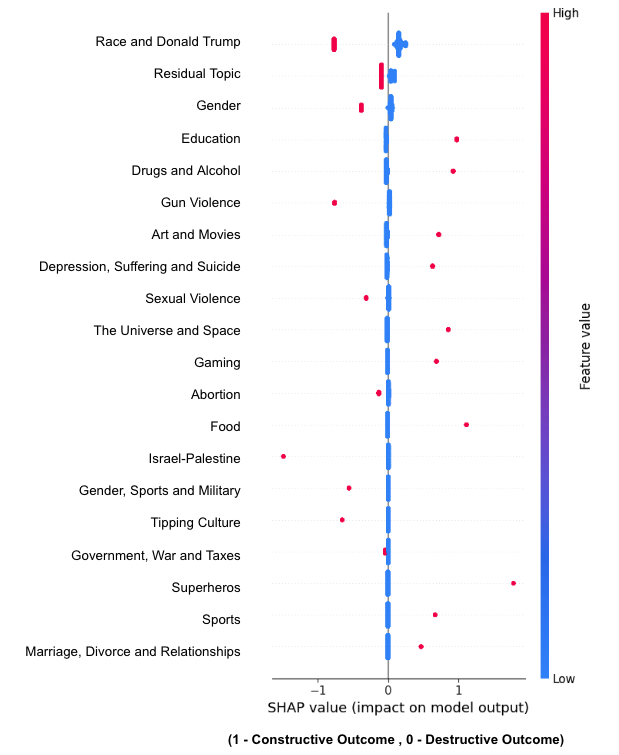} 
    \caption{Reddit Data: Content (Topic Only)} 
    \label{fig:real_3} 
\end{figure}

\clearpage 

\subsubsection*{(Model 4) All Content (Semantic + Topic)}
Top 5 features by SHAPley values:
\begin{itemize}
    \item Variance in Discursive Diversity (-)
    \item Discursive Diversity (-)
    \item Mimicry - BERT (-)
    \item LIWC - Relative (+)
    \item Within Person Discursive Diversity Range (-)
\end{itemize}

Here, the results are largely identical with those of Model 1.

\begin{figure}[ht]
    \centering
    \includegraphics[width=0.9\textwidth,height=0.5\textheight]{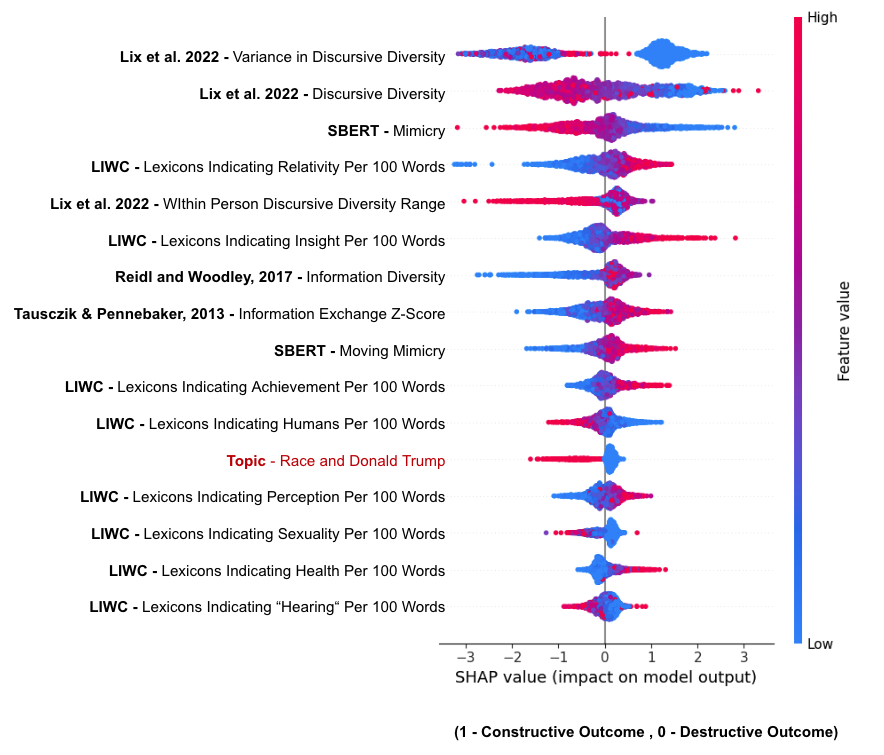} 
    \caption{Reddit Data: All Content (Topic + Semantic)} 
    \label{fig:real_4}
\end{figure}

\clearpage 

\subsubsection*{(Model 5) All TCT (Expression + Semantic)}
Top 5 features by SHAPley values:
\begin{itemize}
    \item Turn-Taking Index (-)
    \item Variance in Discursive Diversity (-)
    \item RoBERTa - Negative (-)
    \item Gini Coefficient - Sum of Number of Messages (-)
    \item Discursive Diversity (-)
\end{itemize}

Here, the results capture a combination of the patterns observed in Models 1 and 2-4; as in Model 1, a high degree of turn-taking, a more unequal conversation (in which fewer individuals dominate the conversation), and more negativity are associated with more destructive outcomes. As in Models 2-4, Variance in Discursive Diversity and Discursive Diversity are also both associated with destructive outcomes.

\begin{figure}[ht]
    \centering
    \includegraphics[width=0.9\textwidth,height=0.5\textheight]{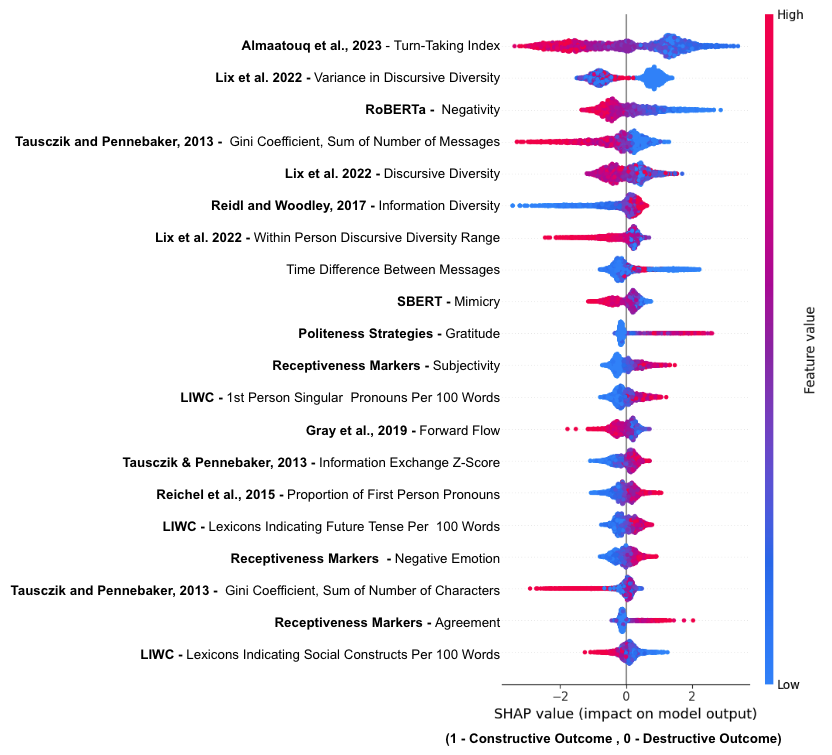} 
    \caption{Reddit Data: All Team Communication Toolkit Features} 
    \label{fig:real_5}
\end{figure}

\clearpage 

\subsubsection*{(Model 6) All Features}
Top 5 features by SHAPley values:
\begin{itemize}
    \item Turn-Taking Index (-)
    \item Variance in Discursive Diversity (-)
    \item Gini Coefficient - Sum of Number of Messages (-)
    \item Discursive Diversity (-)
    \item RoBERTa - Negative (-)
\end{itemize}

Here, the results are largely similar to those of Model 5.

\begin{figure}[ht]
    \centering
    \includegraphics[width=0.9\textwidth,height=0.5\textheight]{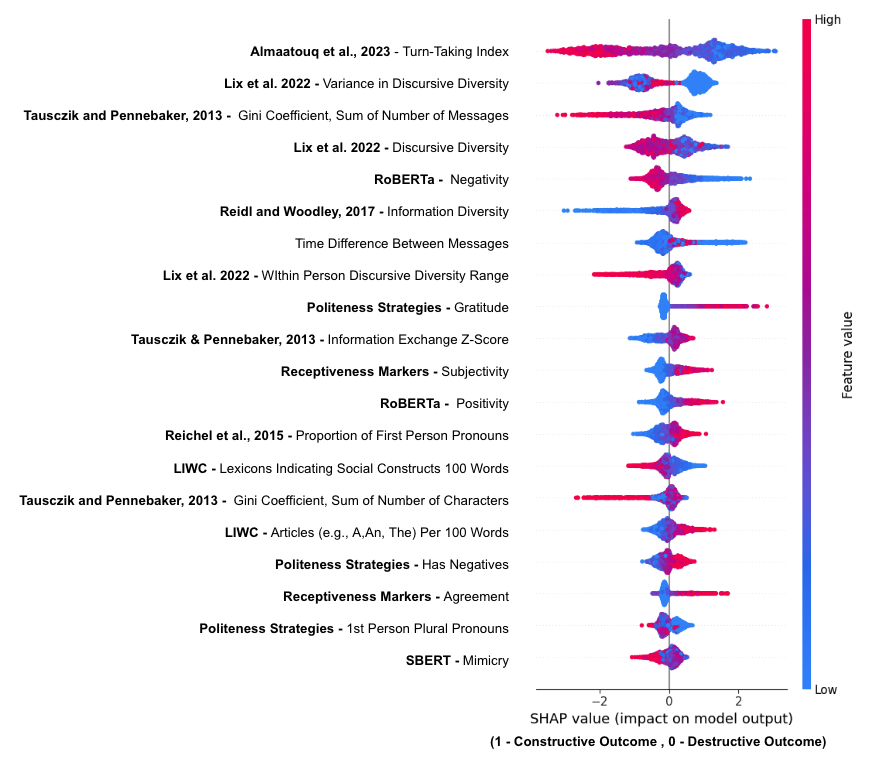} 
    \caption{Reddit Data: All Features} 
    \label{fig:real_6} 
\end{figure}
\clearpage 

\subsection{Team Communication Toolkit Features}~\label{tct}

Table~\ref{tct-features} lists the 150 features we extract for a given \textit{utterance} by the Team Communication Toolkit~\cite{Hu2024}. The features are grouped by a high-level conceptual category. In our analysis, conducted at the level of a conversation, utterance-level features are aggregated by their mean value across the conversation.

\begingroup\tiny
\begin{longtable}{>{\raggedright\arraybackslash}p{1.5cm} >{\raggedright\arraybackslash}p{2.5cm} >{\raggedright\arraybackslash}p{5cm} >{\raggedright\arraybackslash}p{3cm} >{\centering\arraybackslash}p{1cm}}
\caption{Utterance Level Features} \\
\hline
\textbf{Feature Category} & \textbf{Feature(s)} & \textbf{Definition} & \textbf{Citation} & \textbf{\# Features} \\
\hline
\endfirsthead
\caption[]{(continued)} \\
\hline
\textbf{Behavioral Feature (Group)} & \textbf{Computational Feature(s)} & \textbf{Definition} & \textbf{Citation} & \textbf{\# Features} \\
\hline
\endhead

\hline \multicolumn{4}{r}{\textit{Continued on next page}} \\
\endfoot

\hline
\endlastfoot

\textbf{Quantity} & Number of Messages & The number of messages (instances in which the user presses ‘Enter’ and records a distinct utterance. In a turn-based preprocessing option, we combine consecutive messages sent by the same person as a single “turn.” & (Cao et al. 2021); (Marlow et al. 2018), as objective communication frequency; and (Yeomans et al. 2023) for combining utterances into turns. & 1 \\

& Number of Words & The number of space-delimited words. & (Ranganath, Jurafsky, and McFarland 2013); (Cao et al. 2021) & 1 \\

& Number of Characters & The number of alphanumeric characters. & Original Feature & 1 \\

\textbf{Content} & Information Exchange Z-Score (chats) & “Information exchange” is defined as the total word count minus first-person singular pronouns. This value is then z-scored, both within each conversation (e.g., “greatest information exchange within this discussion”) and across conversations (e.g., “greatest information exchange across all discussions). & (Tausczik and Pennebaker 2013) & 1 \\

& Proportion of First Person Pronouns & The total number of First-Person Pronouns by the total number of words. & (Reichel et al. 2015) & 1 \\

& Word Type-Token Ratio & The total number of unique words divided by the total number of words. & (Reichel et al. 2015) & 1 \\

& TextBlob Subjectivity & The score, from [0.0, 1.0], of how “objective” versus “subjective” a statement is, as calculated by TextBlob. & (Cao et al. 2021) & 1 \\

& Online Discussion Tags & Calculates a number of metrics specific to communications in an online setting: 
\begin{itemize}[nosep, left=0pt, label=--]
    \item Num all caps: Number of words that are in all caps 
    \item Num links: Number of links to external resources 
    \item Num Reddit Users: Number of usernames referred to, in u/RedditUser format. 
    \item Num Emphasis: The number of times someone used **emphasis** in their message 
    \item Num Bullet Points: The number of bullet points used in a message. 
    \item Num Line Breaks: The number of line breaks in a message. 
    \item Num Quotes: The number of “quotes” in a message. 
    \item Num Block Quotes Responses: The number of times someone uses a block quote (“>”), indicating a longer quotation 
    \item Num Ellipses: The number of times someone uses ellipses (…) in their message 
    \item Num Parentheses: The number of sets of fully closed parenthetical statements in a message 
    \item Num Emoji: The number of emoticons in a message, e.g., “:)”
\end{itemize} & New & 12 \\

& LIWC Content Features (e.g., Cognitive, Perceptual, Social, Biological, Personal) & The counts of the LIWC lexicons, normalized per 100 words (as recommended by Yeomans et al., 2023). & (Niederhoffer and Pennebaker 2002); (Pennebaker, Mayne, and Francis 1997); (Tausczik and Pennebaker 2010) & 55 \\

& Dale-Chall Readability Score & The Dale-Chall readability score: 0.1579 (proportion of difficult words) + 0.0496 (average sentence length) “Easy” words are determined by the 3,000-word list; unlisted words are difficult. & (Cao et al. 2021) & 1 \\

\textbf{Engagement} & Politeness Features & The politeness discourse markers from ConvoKit’s politenessStrategies. & (Danescu-Niculescu-Mizil et al. 2013) & 21 \\

& Hedging Features & The count of hedge words (“sort of,” “kind of,” “I guess,” “I think,” “a little,” “maybe,” “possibly,” and “probably”), normalized per 100 words. Another version of the feature is simply binary: 1 if it contains any hedge words; 0 otherwise. ConvoKit also provides native hedging features. & (Ranganath, Jurafsky, and McFarland 2013; Danescu-Niculescu-Mizil et al. 2013; Islam, Xiao, and Mercer 2020) & 1 \\

& Receptiveness Markers & A collection of conversational markers indicating the use of politeness / receptiveness. & (Yeomans et al., 2020) & 39 \\

& Conversational Repair & A binary indicator for whether the user attempted to repair the conversation. Calculated using the following regular expression: \texttt{what\textbackslash?+|sorry|excuse me|huh\textbackslash??|}\newline\texttt{who\textbackslash?+|pardon\textbackslash?+|say.*again\textbackslash??|what'?s that|what is that}. & (Ranganath, Jurafsky, and McFarland 2013) & 1 \\

& Number of Questions & The naive version of the feature simply counts the number of question marks. ConvoKit implements a less naive version, a binary indicator for whether there is a “direct question.” & (Ranganath, Jurafsky, and McFarland 2013) & 1 \\

\textbf{Variance} & Function Word Accommodation & The number of function words from a given turn that also appear in the previous turn. Function words were defined by the list in Ranganath et al. (2013). & (Ranganath, Jurafsky, and McFarland 2013) & 1 \\

& Content Word Accommodation & The number of content words from a given turn that also appear in the previous turn, normalized by the term frequency of the content word across all chats in a given dataset. Content words are defined as all words not present on the function word list. & (Ranganath, Jurafsky, and McFarland 2013) & 1 \\

& BERT (Mimicry) & The cosine similarity of the SBERT vectors between the current utterance and the utterance in the previous turn. & (Matarazzo and Wiens 1977); language style matching (Tausczik and Pennebaker 2013); synchrony (Niederhoffer and Pennebaker 2002); implemented in a manner similar to forward flow (Gray et al. 2019) & 1 \\

& Moving Mimicry & The running average of all BERT Mimicry scores computed so far in a conversation. Captures the extent to which all participants in a conversation mimic each other up until a given point. & (Matarazzo and Wiens 1977); language style matching (Tausczik and Pennebaker 2013); synchrony (Niederhoffer and Pennebaker 2002); implemented in a manner similar to forward flow (Gray et al. 2019) & 1 \\

& Forward Flow & The extent to which a conversation “flows forward” — that is, evolves to new topics over time. The forward flow of a given message is the cosine similarity between the SBERT vector of the current message and the average SBERT vector of all previous messages. In other words, it captures how similar a message is to everything that has come before (so far). & (Gray et al., 2019) & 1 \\

\textbf{Pace} & Burstiness & The level of burstiness of chats in a conversation. The metric takes a value between -1 and 1, with a higher value indicating higher levels of team burstiness. Teams with higher burstiness would have more spiked patterns in team activity, which tends to indicate a higher sense of responsiveness and connectedness within the team members. & (Reidl and Woolley, 2017) & 1 \\

& Time Difference Between Messages & The difference in timestamp between subsequent messages. & (Reichel et al. 2015) & 1 \\

\textbf{Emotion} & BERT Sentiment (positive, negative, and neutral) & The values for the labels “positive,” “negative,” and “neutral” from the pre-trained model. & (“Cardiffnlp/twitter-Roberta-Base-Sentiment-Latest · Hugging Face,” n.d.) & 3 \\

& Positivity Z-Score (within- and across-conversation) & The z-score of how “positive” a chat is, relative to other chats (1) in the same conversation. & (Tausczik and Pennebaker 2013) & 1 \\

& TextBlob Polarity & The score, from [-1.0, 1.0], of how “positive” or “negative” a statement is, as calculated by TextBlob. & (Cao et al. 2021) & 1 \\

\hline
\multicolumn{4}{l}{\textbf{Total Number of Utterance-Level Features}} & \textbf{150}
\label{tct-features}
\end{longtable}
\endgroup

Table~\ref{tct-conversation-features} lists the 10 features we extract at the level of an entire \textit{conversation}; that is, rather than describing an attribute of a specific utterance, these features describe a feature of the overall collection of all utterances --- for example, whether participants in the conversation shared airtime equally, whether participants made semantically similar points; and whether participants tended to speak in short temporal ``bursts,'' as opposed to communicating at a constant rate throughout the conversation.

\begingroup\tiny
\begin{longtable}{>{\raggedright\arraybackslash}p{1.5cm} >{\raggedright\arraybackslash}p{2.5cm} >{\raggedright\arraybackslash}p{5cm} >{\raggedright\arraybackslash}p{3cm} >{\centering\arraybackslash}p{1cm}}
\caption{Conversation Level Features} \\
\hline
\textbf{Feature Category} & \textbf{Feature(s)} & \textbf{Definition} & \textbf{Citation} & \textbf{\# Features} \\
\hline
\endfirsthead
\caption[]{(continued)} \\
\hline
\textbf{Feature Category} & \textbf{Feature(s)} & \textbf{Definition} & \textbf{Citation} & \textbf{\# Features} \\
\hline
\endhead
\hline \multicolumn{4}{r}{\textit{Continued on next page}} \\
\endfoot
\hline 
\endlastfoot
\textbf{Equality} & Equal Participation & The extent to which each participant in a conversation engages equally, as measured by a Gini coefficient. We calculate three flavors of Gini coefficient, using the number of words, number of characters, and the number of messages, respectively. & (Tausczik \& Pennebaker, 2013) & 3 \\

& Turn-Taking Index & Calculates a metric describing the extent to which individuals take turns speaking in a conversation. Adapted from Almaatouq et al. (2023), in which we treat each separate chat as equivalent to an in-game “solution”: ”A group’s turn-taking index for a given round is measured by dividing the number of turns taken … by the total number of [chats] on a particular task instance.” & (Almaatouq et al., 2023) & 1 \\

\textbf{Variance} & Discursive Diversity & Calculates metrics related to the extent to which members in a conversation speak similarly. 
\begin{itemize}[nosep, left=0pt, label=--]
    \item Discursive diversity: 1 - the average pairwise cosine distances between the centroids associated with each speaker in a conversation. 
    \item Variance in discursive diversity: the extent to which discursive diversity varies across the beginning, middle, and end of a conversation. 
    \item Incongruent modulation: the total variance, per speaker, between the (beginning, middle) and (middle, end) of a conversation. As described by the pape, this is the “team-level variance in members’ within-person discursive range” from stage 1 to stage 2, and from stage 2 to stage 3. 
    \item Within-person discursive range: The sum, across all speakers in the conversation, of each speaker’s average distance between their centroids for the (beginning, middle) and (middle, end) of a conversation.
\end{itemize} & (Lix et al., 2022) & 4 \\

& Information Diversity & This conversation-level feature uses topic modeling to measure the level of information diversity across a conversation. We first preprocess the data with lowercasing, lemmatization, removing stop words, and removing short words (less than length 3). We then use the gensim package to create an LDA Model for each conversation, generating a corresponding topic space with its number of dimensions = num\_topics. To determine the number of topics used, we use a logarithmic scale relative to the number of chats in the conversation. A team's info diversity is then computed by looking at the average cosine dissimilarity between each chat's topic vector and the mean topic vector across the entire conversation. & (Reidl and Woolley, 2017) & 1 \\

\textbf{Pace} & Burstiness & This conversation-level feature measures the level of burstiness of chats in a conversation. The metric takes a value between -1 and 1, with a higher value indicating higher levels of team burstiness. Teams with higher burstiness would have more spiked patterns in team activity, which tends to indicate a higher sense of responsiveness and connectedness within the team members. & (Reidl and Woolley, 2017) & 1 \\
\hline
\multicolumn{4}{l}{\textbf{Total Number of Conversation Level Features}} & \textbf{10}
\label{tct-conversation-features}
\end{longtable}
\endgroup

\subsection{Classification of Features as ``Expression'' or ``Content-Semantic''}

The first and last authors classified the features from the Team Communication Toolkit package as related to "Content-Semantic" and "Expression" based on a general rule of thumb that if a speaker could rephrase their message to avoid triggering a particular feature without changing the meaning, the feature describes the \textit{expression}, whereas features that pertain to the message's inherent meaning are \textit{content} features.

Specific examples include the following:

\begin{itemize}
    \item \textbf{\textit{Basic grammar elements}} (e.g., pronouns, verbs) are considered expression features, as they dictate sentence structure without changing the core meaning.
    
    \item \textbf{\textit{Stopwords}} are categorized as expression, as they can often be omitted without altering the intent of a message.

    \item \textbf{\textit{Questions}} are classified as expression because the same idea can be framed as a question or a statement, with the question form often playing a significant role in Directness.

    \item \textbf{\textit{Turn-Taking, Equality, and other dynamics of conversational flow}} are classified as expression because they describe speaker interactions that do not affect the meaning of the conversation.
\end{itemize}

Based on these rules, the following is how we categorized features from the Team Communication Toolkit into ``Expression'' and ``Content-Semantic'' as follows. The feature names listed here are the names of the columns as returned by the Team Communication Toolkit.

\subsubsection*{Expression Related Features : Utterance Level}
\begingroup\tiny
\begin{multicols}{3}
\begin{itemize}[leftmargin=*,noitemsep]
    \item positive\_bert
    \item negative\_bert
    \item discrepancies\_lexical\_per\_100
    \item neutral\_bert
    \item num\_words
    \item num\_chars
    \item num\_messages
    \item conjunction\_lexical\_per\_100
    \item certainty\_lexical\_per\_100
    \item inclusive\_lexical\_per\_100
    \item bio\_lexical\_per\_100
    \item adverbs\_lexical\_per\_100
    \item third\_person\_lexical\_per\_100
    \item negation\_lexical\_per\_100
    \item swear\_lexical\_per\_100
    \item negative\_affect\_lexical\_per\_100
    \item quantifier\_lexical\_per\_100
    \item positive\_affect\_lexical\_per\_100
    \item present\_tense\_lexical\_per\_100
    \item future\_tense\_lexical\_per\_100
    \item past\_tense\_lexical\_per\_100
    \item inhibition\_lexical\_per\_100
    \item sadness\_lexical\_per\_100
    \item social\_lexical\_per\_100
    \item indefinite\_pronoun\_lexical\_per\_100
    \item anger\_lexical\_per\_100
    \item first\_person\_singular\_lexical\_per\_100
    \item feel\_lexical\_per\_100
    \item tentativeness\_lexical\_per\_100
    \item exclusive\_lexical\_per\_100
    \item verbs\_lexical\_per\_100
    \item article\_lexical\_per\_100
    \item argue\_lexical\_per\_100
    \item auxiliary\_verbs\_lexical\_per\_100
    \item cognitive\_mech\_lexical\_per\_100
    \item preposition\_lexical\_per\_100
    \item first\_person\_plural\_lexical\_per\_100
    \item second\_person\_lexical\_per\_100
    \item positive\_words\_lexical\_per\_100
    \item first\_person\_lexical\_per\_100
    \item nltk\_english\_stopwords\_lexical\_per\_100
    \item hedge\_words\_lexical\_per\_100
    \item num\_question\_naive
    \item NTRI
    \item word\_TTR
    \item first\_pronouns\_proportion
    \item function\_word\_accommodation
    \item hedge\_naive
    \item textblob\_subjectivity
    \item textblob\_polarity
    \item positivity\_zscore\_chats
    \item dale\_chall\_score
    \item time\_diff
    \item please
    \item please\_start
    \item hashedge
    \item indirect\_btw
    \item hedges
    \item factuality
    \item deference
    \item gratitude
    \item apologizing
    \item 1st\_person\_pl
    \item 1st\_person
    \item 1st\_person\_start
    \item 2nd\_person
    \item 2nd\_person\_start
    \item indirect\_greeting
    \item direct\_question
    \item direct\_start
    \item haspositive
    \item hasnegative
    \item subjunctive
    \item indicative
    \item Acknowledgement
    \item Affirmation
    \item Agreement
    \item Apology
    \item Ask\_Agency
    \item By\_The\_Way
    \item Can\_You
    \item Conjunction\_Start
    \item Could\_You
    \item Disagreement
    \item Filler\_Pause
    \item First\_Person\_Plural
    \item First\_Person\_Single
    \item For\_Me
    \item For\_You
    \item Formal\_Title
    \item Give\_Agency
    \item Goodbye
    \item Gratitude
    \item Hedges
    \item Hello
    \item Impersonal\_Pronoun
    \item Informal\_Title
    \item Let\_Me\_Know
    \item Negation
    \item Negative\_Emotion
    \item Please
    \item Positive\_Emotion
    \item Reasoning
    \item Reassurance
    \item Second\_Person
    \item Subjectivity
    \item Swearing
    \item Truth\_Intensifier
    \item Bare\_Command
    \item YesNo\_Questions
    \item WH\_Questions
    \item Adverb\_Limiter
    \item Token\_count
    \item certainty\_rocklage
    \item num\_all\_caps
    \item num\_reddit\_users
    \item num\_emphasis
    \item num\_bullet\_points
    \item num\_numbered\_points
    \item num\_line\_breaks
    \item num\_ellipses
    \item num\_parentheses
    \item num\_emoji
\end{itemize}
\end{multicols}
\endgroup

\subsubsection*{Expression Related Features : Conversation Level}
\begingroup\tiny
\begin{itemize}[leftmargin=*,noitemsep]
    \item turn\_taking\_index
    \item gini\_coefficient\_sum\_num\_words
    \item gini\_coefficient\_sum\_num\_chars
    \item gini\_coefficient\_sum\_num\_messages
    \item team\_burstiness
\end{itemize}
\endgroup

\subsubsection*{Content Related Features: Utterance Level}
\begingroup\tiny
\begin{multicols}{2}
\begin{itemize}[leftmargin=*,noitemsep]
    \item info\_exchange\_zscore\_chats
    \item hear\_lexical\_per\_100
    \item home\_lexical\_per\_100
    \item achievement\_lexical\_per\_100
    \item anxiety\_lexical\_per\_100
    \item death\_lexical\_per\_100
    \item health\_lexical\_per\_100
    \item see\_lexical\_per\_100
    \item body\_lexical\_per\_100
    \item family\_lexical\_per\_100
    \item insight\_lexical\_per\_100
    \item humans\_lexical\_per\_100
    \item relative\_lexical\_per\_100
    \item sexual\_lexical\_per\_100
    \item religion\_lexical\_per\_100
    \item work\_lexical\_per\_100
    \item money\_lexical\_per\_100
    \item causation\_lexical\_per\_100
    \item friends\_lexical\_per\_100
    \item percept\_lexical\_per\_100
    \item num\_links
    \item num\_quotes
    \item num\_block\_quote\_responses
    \item mimicry\_bert
    \item moving\_mimicry
    \item forward\_flow
    \item content\_word\_accommodation
\end{itemize}
\end{multicols}
\endgroup

\subsubsection*{Content Related Features: Conversation Level}
\begingroup\tiny
\begin{itemize}[leftmargin=*,noitemsep]
    \item info\_diversity
    \item discursive\_diversity
    \item variance\_in\_DD
    \item incongruent\_modulation
    \item within\_person\_disc\_range
\end{itemize}
\endgroup

\subsection{Data and Cleaning}~\label{datasets}

\subsubsection{Example Data} \label{sample-data}
Table~\ref{sample-conversation-table} shows four example conversations from our data. The top row shows two conversations in which the outcome is ``destructive,'' and the bottom row shows two conversations in which the outcome is ``constructive.'' The first column shows examples from the Synthetic dataset, and the second column shows examples from the Reddit dataset (all speaker identifiers have been transformed to a standard format).

\begin{table}[ht]
\centering
\begingroup\tiny
\begin{tabular}{>{\raggedright\arraybackslash}m{6cm} >{\raggedright\arraybackslash}m{6cm}}
\hline
\multicolumn{2}{c}{\textbf{Destructive Conflict}} \\
\hline
\textbf{Example Synthetic Conversation} & \textbf{Example Reddit Conversation} \\
\hline
\textbf{speaker\_1}: Why didn't you tell me you were going to change the project topic? & \textbf{speaker\_1}: You should not rest your feet on something that is used by everyone for sitting. Feet/shoes are considered to be dirty and rightly so. \\
\textbf{speaker\_2}: Because its none of your business. I can choose whatever topic I want. & \textbf{speaker\_2}: I don't believe you're correct, I don't consider my shoes to be any more dirty than a chair. \\
\textbf{speaker\_1}: Of course it's my business! We're supposed to be working on this together! & \textbf{speaker\_3}: I don't consider my saliva to be any dirtier than my feet. Should I be able to spit on the floor indoors with no consequences? \\
\textbf{speaker\_2}: Well, maybe I got tired of doing all the work by myself! & \textbf{speaker\_2}: No, you're mistaken. Saliva is dirtier than your feet (hopefully). \\
\textbf{speaker\_1}: Excuse me? I've been contributing just as much as you have! & \textbf{speaker\_4}: So, I'm seeing you contradict yourself. You say saliva is definitely dirtier than your shoes, why? If dirtier is subjective to visible crud, then why is saliva definitely dirtier than your shoes? Because of germs? Do you think that germs don't live outside your mouth? Or just more dangerous ones are in the mouth so visible crud doesn't apply here? \\
\textbf{speaker\_2}: Really? Because I don't see any of your work in the shared folder. & \\
\textbf{speaker\_1}: Maybe if you checked your emails, you'd see that I've been sending you updates. & \\
\textbf{speaker\_2}: How am I supposed to know that if you don't tell me? Mind reading? & \\
\hline
\multicolumn{2}{c}{\textbf{Constructive Conflict}} \\
\hline
\textbf{Example Synthetic Conversation} & \textbf{Example Reddit Conversation} \\
\hline
\textbf{speaker\_1}: I can't believe you pulled the goalie without telling anyone! & \textbf{speaker\_1}: What's your basis for believing that SuperPACs make elections un-free or un-fair? \\
\textbf{speaker\_2}: The game was slipping away, I had to make a quick decision. & \textbf{speaker\_2}: Well in my brain a fair election is when anyone can run for office. The fact that you need millions of dollars to do so seems unfair. It instantly results in a situation where only the rich can run for and hold offices. On the municipal level I don't think superpacs are a problem, but definitely on a federal level. \\
\textbf{speaker\_1}: But we had no plan! It threw everyone off! & \textbf{speaker\_1}: Anyone \textit{can} run for office. You might not get very far, but you can run. And you don't need millions of dollars, you need to be \textit{able to raise} millions of dollars. In itself, I don't see a problem with that. Raising large amounts of money requires leadership, diplomacy, organization, charisma — all traits I like to see in my elected officials. The corruption and lack of transparency created by PACs and SuperPACs and Citizens United in general is definitely a major problem, but like some other things that have come up in this discussion, it's not an issue that makes the USA an un-democratic state. There's no rule saying that the person who raises the most money wins the election. The fact that most people vote for the most visible candidate on their side rather than the one that most closely represents their interests is a problem with the electorate and also a result of our winner-take-all voting system, but neither of those are in conflict with the USA being a democracy by definition. \\
\textbf{speaker\_2}: I understand it was unexpected. I should have communicated better. & \textbf{speaker\_2}: Okay, I think you have officially Changed My View haha... it seems like the problem is more caused by the populace not being responsible voters, I see that and agree with it. Ideally I would like it if no money was involved in the political process, but that is naive. \\
\textbf{speaker\_1}: Yeah, exactly. We need to be on the same page in those crucial moments. & \\
\textbf{speaker\_2}: You're right. Next time, I'll make sure to signal the change and discuss it first. & \\
\hline
\end{tabular}
\endgroup
\vspace{0.5cm}
\centering
\caption{Excerpts from conversations in our data. Our data comprises two sources: synthetic conversations generated by GPT-4 (left column), and Reddit conversations from r/ChangeMyView (right column). The outcomes of the conversations are either destructive (top row) or constructive (bottom row).}
\label{sample-conversation-table}
\end{table}

\subsubsection{Synthetic Data: Prompts for GPT-4}~\label{gpt-prompts}

We generated the Synthetic data using the following prompts and the GPT-4 API (``gpt-4-1106-preview'').

\subsubsection*{Synthetic Utterances (Used for Validating Measures of Conflict Expression)}

For each utterance generated, we provided a system prompt with context about Directness and Oppositional Intensity, alongside specific examples of the constructs:

\begin{quote}

The most important thing about the message you generate is that it varies from the previous messages in language and structure. You should minimize the amount of repeated phrases you are using. \\

\textit{\textbf{Summary of Directness and Oppositional Intensity}}

Here, we summarize directness and oppositional intensity to help refresh you on what to look for. 
Please use this section as a handy reference while you work on the labeling task.

\textit{\textbf{Directness:}}
The directness of an expression refers to the degree to which the sender explicitly conveys their opposition 
to another person. A person is direct when they are clear and explicit with their opinion, leaving little room 
for confusion or misinterpretation.

\textit{\textbf{Examples of direct vs. indirect expressions:}}

\textit{\textbf{Use of Ambiguity:}}
Direct: "I disagree with that proposal."
Indirect: "I see your point, but I might have a slightly different perspective on the proposal."
Indirect: Storytelling, reflective questions
E.g., What information would make you change your mind?

\textbf{\textit{Use of Certainty:}}
Direct: "No, I can't attend the meeting tomorrow."
Indirect: "I'm not sure if I'll be available for the meeting tomorrow."

\textbf{\textit{Use of Sarcasm:}}
Direct: "Wow, I think that presentation went very badly."
Indirect: "[sarcastic] Well, that presentation certainly sets a bar, doesn't it?"

Some points to consider while rating directness:
- Is the speaker trying to hedge or avoid certain issues? (Hedging is indirect.)
- Is sarcasm present? (Sarcasm is indirect.)
- Is the message unclear? (Unclear communication is indirect.)
- Is the speaker expressing uncertainty? (Uncertainty is indirect.)

\textbf{\textit{Oppositional Intensity:}}
Oppositional Intensity refers to the degree of force with which opposition is conveyed. High-intensity conflict 
is conveyed with greater force, while lower-intensity conflicts are more measured.

\textbf{\textit{Key concepts: Entrenchment and Subversion}}
Entrenchment: A person "digs in" their position, refusing to listen to other ideas.
Subversion: A person undermines others, potentially through personal attacks or attempts to block their voice.

\textit{\textbf{Points to consider for oppositional intensity:}}
- Is the disagreement expressed with strong energy or forcefulness?
- Does the speaker show emotional intensity?
- Is there evidence of entrenchment, meaning an unwillingness to hear the other side?
- Is subversive behavior present (e.g., personal attacks, blocking others)?

\textbf{\textit{Examples of oppositional intensity:}}

\textbf{\textit{Emotional Activation:}}
High: "I am incredibly angry about this new law. It goes against everything I believe in and I hate the people 
who put it through Congress."
Low: "This new law makes me sad. It's really disappointing to see the country moving in this direction."

\textbf{\textit{Use of Entrenchment:}}
High: "This is the entirely wrong decision. There is no possible justification for it."
Low: "I don’t agree with this decision, but I’d be curious to hear their reasoning."

\textbf{\textit{Subversive Behavior:}}
High: "Shut up, you stupid b*tch."
Low: "OK, you do you."
\end{quote}

We then prompted GPT-4 (``gpt-4-1106-preview'') to generate individual messages that were either high or low in each of Directness and Oppositional Intensity:

\begin{quote}
    Please generate sentences with {HIGH/LOW} directness and {HIGH/LOW} oppositional intensity.
\end{quote}

\subsubsection*{Synthetic Conversations (Used to Generate Main Synthetic Dataset)}

To generate conversations that either escalated to a destructive outcome or de-escalated to a constructive outcome, we used the following prompt:

\begin{quote}
For the following messages you provide for the conflict dialogue, I want them to be in the form “speakerX: message”. Every line should have a colon and X should always be a number. Avoid giving stage directions or emotional states.
I want to simulate a {0} conflict conversation between two or more parties. These parties should be related in some way and have a particular dynamic (friends arguing, parent scolding child, sports team fighting when losing, teacher confronting a student about cheating, etc.). Your job is to stimulate this conflict through conversation. Make sure that the parties involved in the conversation are over conflict about the same task, process, or outcome. Also, try to be unique with the scenario.
Every line must follow the format “speakerX: message”. Do not include any lines without a colon.

Escalatory conflict spirals are exchanges characterized by reciprocated negative communications, such as threats or other tactics that suppress information availability and usage, that are difficult to break and generally produce negative outcomes for the participants. De-escalatory conflict spirals are exchanges characterized by reciprocated information exchange and complementary questioning and answering that generally produce positive outcomes for participants.
\end{quote}

\subsubsection*{Generation of Metadata}

After receiving the generated text from GPT-4, we computationally added meta-data such as the speaker identifier, conversation identifier, and the timestamp as arbitrary values, as these are required for running the Team Communication Toolkit. Sample code for generating metadata for Synthetic Utterances is provided below.

\begingroup\scriptsize
\begin{verbatim}
conversation_num = 1 
with open('generated_sentences.csv', 'w', newline='') as csvfile:
    fieldnames = ['conversation_num', 'speaker_id', 'message', 'timestamp', 
        'directness', 'oppositional_intensity'] 
    writer = csv.DictWriter(csvfile, fieldnames=fieldnames)

    writer.writeheader()

    for description, (directness, oppositional_intensity) in categories.items():
        print(f"\nGenerating sentences for: {description}\n")
        
        # Custom prompt for each category
        category_prompt = system_prompt + f"\n\nPlease generate sentences with {directness} directness 
            and {oppositional_intensity} oppositional intensity."
        
        sentences = generate_sentences(category_prompt)
        for idx, sentence in enumerate(sentences):
            writer.writerow({
                'conversation_num' : conversation_num, 
                'speaker_id' :  f"Speaker_{random.randint(0,5)}",
                'message' : sentence,
                'timestamp' : datetime.now().strftime('%Y-%m-%d %H:%M:%S'),
                'directness' : directness,
                'oppositional_intensity' : oppositional_intensity

            })
            conversation_num += 1 

        print(f"Sentences for {description} saved to CSV.")
\end{verbatim}
\endgroup

\subsubsection{Reddit Data: Cleaning and Pre-processing} \label{reddit-data-processing}
The original datasets consist of a total of 15,360 conversations (120,894 utterances) from “Conversations Gone Awry”~\cite{Chang2019}  and “Winning Conversations”~\cite{Tan2016} The posts from these datasets originate from the subreddit r/ChangeMyView, and were made on Reddit between 2013 and 2018. The following fields are present in the respective datasets:

\subsubsection*{Columns Present: Conversations Gone Awry Dataset}
Each utterance corresponds to a Reddit comment and involves the following columns:
\begin{itemize}
    \item \textbf{id}: Reddit ID of the comment represented by the utterance
    \item \textbf{speaker}: the speaker who authored the utterance
    \item \textbf{conversation\_id}: id of the first utterance in the conversation this utterance belongs to. Note that this differs from how ‘conversation\_id’ is treated in ConvoKit’s general Reddit corpora: in those corpora a conversation is considered to start with a Reddit post utterance, whereas in this corpus a conversation is considered to start with a top-level reply to a post.
    \item \textbf{reply\_to}: Reddit ID of the utterance to which this utterance replies to (None if the utterance represents a top-level comment, i.e., a reply to a post)
    \item \textbf{timestamp}: time of the utterance
    \item \textbf{text}: textual content of the utterance
    \item \textbf{meta.score}: score (i.e., the number of upvotes minus the number of downvotes) of the content
    \item \textbf{meta.top\_level\_comment}: the id of the top level comment (None if the utterance is a post)
    \item \textbf{meta.retrieved\_on}: unix timestamp of the time of when the data is retrieved
    \item \textbf{meta.gilded}: gilded status of the content
    \item \textbf{meta.gildings}: gilding information of the content
    \item \textbf{meta.stickied}: stickied status of the content
    \item \textbf{meta.permalink}: permanent link of the content
    \item \textbf{meta.author\_flair\_text}: flair of the author
\end{itemize}

\subsubsection*{Columns Present: Winning Corpus Dataset}
Each utterance corresponds to a Reddit comment and involves the following columns:
\begin{itemize}
    \item \textbf{id}: index of the utterance (unique comment identification provided by Reddit)
    \item \textbf{speaker}: the unique id of the user who authored the utterance
    \item \textbf{conversation\_id}: comment identifier of the original post in the thread that this comment was posted in
    \item \textbf{reply\_to}: index of the utterance to which this utterance replies to (None if the utterance is not a reply)
    \item \textbf{timestamp}: utterance timestamp provided by Reddit API
    \item \textbf{text}: the full text (in string) of the comment
    \item \textbf{meta.success}: an indicator taking the value of 1 if the comment was part of a successful argument thread (i.e. an argument thread that changed the OP’s mind), 0 if unsuccessful, and None if not part of either a successful or unsuccessful thread.
    \item \textbf{meta.pair\_ids}: every successful-unsuccessful argument pair originally compiled by the authors has a unique pair\_id. However, it is important to note that not every argument is unique (i.e. a single negative argument within a conversation could have two opposing positive arguments, which necessitates two corresponding pair\_ids. Therefore, pair\_ids is a list).
    \item \textbf{meta.replies}: a list of comment ids that respond directly to the current comment. For the OP post in the thread, this was constructed by selecting all comment ids with a “reply\_to” field equal to the original post id (this was necessary because the original data provided by the authors did not include all the children of the OP post in their data format). For all comments besides the original post, the “replies” field was originally provided by Reddit API.
\end{itemize}

The following pieces of metadata were originally provided by Reddit API:
\begin{itemize}
    \item author\_flair\_text, author\_flair\_css\_class, banned\_by, controversiality, edited, distinguished, user\_reports, ups, downs, subreddit\_id, subreddit, score\_hidden, score, saved, report\_reasons, mod\_reports, num\_reports, likes, gilded, approved\_by
\end{itemize}
*Fields preceded by “meta” are inherited from the general CMV corpus.

\subsubsection*{Preprocessing: Winning Corpus Dataset}
In the original “Winning Arguments” paper~\cite{Tan2016}, this corpus was used in a paired prediction task predicting for whether a reply thread (starting from a top-level comment in the comment thread) was successful in convincing the original poster. As stated in Section 4 of the original paper, the threads were paired by first selecting a reply thread that wins a $\Delta$ (i.e. was successful in convincing the OP), then paired with an unsuccessful reply thread in the same discussion tree that did not win a $\Delta$ but was the most “similar” in topic, as measured by Jaccard similarity. The corpus exposes these successful-unsuccessful pairs used in the original paper through Conversation and Utterance-level metadata. It additionally provides the other sibling reply threads for context (as part of the full comment thread).

This means that the original Winning Corpus Dataset contains entire conversations, which are composed of multiple threads.

The data contain a column called pair\_ids: every successful-unsuccessful argument pair originally compiled by the authors has a unique pair\_id. However, it is important to note that not every argument is unique (i.e. a single negative argument within a conversation could have two opposing positive arguments, which necessitates two corresponding pair\_ids. Therefore, pair\_ids is a list).

We eliminate all hyperlinks, which are too long and cause certain models like RoBERTa to break and also eliminate quoted text i.e. when a person quotes another person, as both these do not represent the speaker’s original content.

\subsubsection*{Preprocessing: Conversations Gone Awry Dataset}
The Conversations Gone Awry Dataset~\cite{Chang2019} contains only threads that derail into personal attacks. We processed this data as follows, in order to extract conversations in which a ``delta'' had been awarded (which we used as our indicator of a constructive outcome):
\begin{itemize}
    \item Extract pair threads
    \item Split the pair threads into successful and unsuccessful threads based on the “meta.success” column i.e. whether the OP has awarded a delta or not. We now have threads of successful and unsuccessful conversations
    \item Add the relevant OP's post to each thread so that the RAs can have context while rating and assign pseudo conversation IDs to identify each thread.
\end{itemize}

\textit{\textbf{Preprocessing: Original Poster’s Post.}}
The Original Poster (OP)’s post is present only in the Winning Dataset~\cite{Tan2016}, and not in the Conversations Gone Awry~\cite{Chang2019} dataset. To maintain parity, we drop the OP’s post from the Winning dataset while running our models.

\textit{\textbf{Preprocessing: Splitting Long Chats.}}
In order to more adequately capture moment-to-moment expressions of conflict, we broke up extremely long Reddit comments into “utterances” of 50 words. Additionally, we identified instances of quoted text and considered the paragraph immediately following the quote to be a separate utterance (a direct response to the quote), such as in the following example:

\begin{quote}
> these people are by no means extraordinarily smart, talented, or even **particularly business savvy.** \\
Consider that 70\% of rich families lose their wealth by the Second Generation, and 90\% lose it by the third generation.
\end{quote}

\subsubsection*{Data Resampling to Ensure Distributional Balance}
One concern is that the data from “Awry” (destructive) versus “Winning” (constructive) conversations fundamentally vary on attributes other than the focal conversation features; for example, if destructive conversations are much longer than constructive ones, the difference in length could confound our effects. That is, the driving factor for any effect we see may come from the fact that destructive conversations simply come from a different distribution, rather than the fact that people express themselves differently.

To address this possibility, we balanced the Awry and Winning datasets to ensure equal representation of 4 potential confounders: conversation length, number of words per chat, number of characters per chat, and meta score per chat. This step was crucial to mitigate any bias in the analysis.

For each confounder, we compared both datasets, retaining all conversations from the smaller group and randomly sampling an equivalent number from the larger group. We use this approach to produce a balanced dataset, where both Awry and Winning categories contributed an equal number of conversations for each confounder. By addressing these imbalances, we reduce the chances of uneven distributions influencing our results.

\subsection{Human Annotation of Conflict Expressions in Reddit Data (For Validation)}~\label{human-annotation-proc}

We select a total of 121 conversations (929 utterances) through random sampling, which are assigned to the raters for hand-labeling. The following is the distribution of the conversations that were hand-labeled:

\begin{center}
\begin{tabular}{l c c}
\textbf{Particulars} & \textbf{Awry} & \textbf{Winning} \\
\hline
Number of Chats & 570 & 359 \\
Number of Conversations & 58 & 63 \\
\end{tabular}
\end{center}

We recruited three undergraduate research assistants (RAs), who annotated the data in terms of Directness and Oppositional Intensity, separately coding for the Directness/Oppositional Intensity of a statement’s content and its expression. The RAs were selected from a pool of applicants who were asked to read relevant excerpts of Weingart et. al 2015’s paper to understand the concepts of Directness and Oppositional Intensity. They were then asked to complete a sample labeling task for three conversations. The RAs whose labels had the highest Fleiss’ Kappa measured against the “gold standard” ratings created by the primary author team were recruited for the labeling task. Upon discussion with the RAs, we developed a set of rules and examples to assist in the labeling process.

\textit{\textbf{Questions for Directness.}} The RAs were asked the following questions for every utterance while rating the conversation for Directness:
\begin{enumerate}
    \item \textbf{Clearly convey the content of their opinion?} (Directness of Content)
        \begin{itemize}
            \item Yes - Direct Content
            \item Neutral - Content contains no opinion
            \item No - Indirect Content
        \end{itemize}
    \item \textbf{Express their opinion in a tone or manner that is direct and assertive?} (Directness of Expression)
        \begin{itemize}
            \item Yes - Direct Expression
            \item No - Indirect Expression
        \end{itemize}
\end{enumerate}

\textit{\textbf{Questions for Oppositional Intensity.}} The RAs were asked the following questions for every utterance while rating the conversation for Oppositional Intensity:
\begin{enumerate}
    \item \textbf{Defend their own position in opposition to others’ positions?} (Oppositional Intensity in Content)
        \begin{itemize}
            \item Yes – Content opposes someone else
            \item No – Content does not oppose anyone
        \end{itemize}
    \item \textbf{Express their point(s) with high emotional activation or force?} (Oppositional Intensity of Expression)
        \begin{itemize}
            \item Yes – Expression is emotional/forceful
            \item No – Expression is not emotional/forceful
        \end{itemize}
\end{enumerate}

To validate whether the conversational attributes we extracted effectively captured expression as it is perceived by humans, our models used the attributes to predict the two expression labels (Directness of Expression and Oppositional Intensity of Expression).

\subsection{Topic Modeling}~\label{topic-modeling}
To better understand the topics discussed in the Reddit conversations, we use the Python package BERTopic~\cite{Grootendorst2022} to assign each conversation a topic label. At a high level, BERTopic assigns topics by using a pre-trained transformer-based language model to generate embeddings for documents, then reduces their dimensionality. It then clusters documents within the lower-dimensional space, and finally generates a topic representation using a class-based TF-IDF procedure.\\ 

\begin{figure}[ht]
    \centering
    \includegraphics[width=1.0 \textwidth]{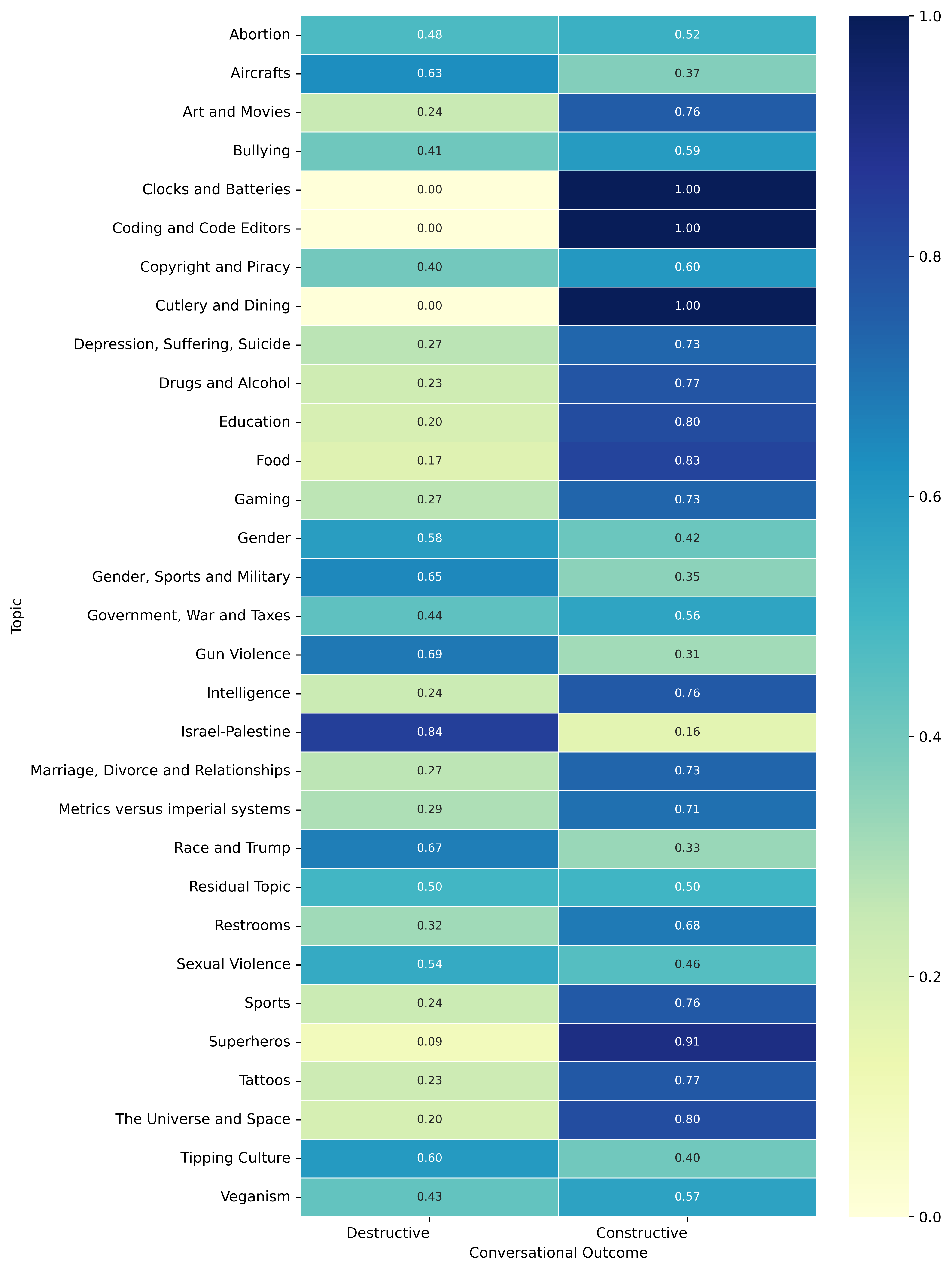} 
    \caption{Distribution of Constructive and Destructive Conversations Across Topics}
    \label{fig:topic_distribution} 
\end{figure}

For the purpose of assigning topic labels, all utterances within the same conversation are concatenated into a single “document,” as we assume that each conversation (a single Reddit thread) is about the same topioc. The output of BERTopic is provided in Table~\ref{table:topic_representation}; after experimenting with 30, 60, and 100 topics (a hyperparameter of of BERTopic), we found that 30 topics captures approximately $65\%$ of the conversations, and increasing the number of topics did not increase the coverage. We therefore classified the conversations into $30$ topics, with the remaining $35\%$ of topics assigned to the ``Residual Topic'' (Row 1 of Table~\ref{table:topic_representation}). Conversations in the Residual Topic equally represented those with constructive and destructive outcomes (Figure~\ref{fig:topic_distribution}). 

For interpretability, we manually assign a human-readable label to these topics based on the representative words generated by BERTopic. For example, the representative words ['child', 'abortion', 'fetus', 'life', 'parents', 'children', 'dont', 'mother', 'people', 'pregnancy'] are assigned the topic of ``Abortion.''

\begin{table}[ht]
\centering
\begingroup\tiny
\begin{tabular}{|>{\raggedleft\arraybackslash}p{8cm}|>{\centering\arraybackslash}p{3cm}|>{\centering\arraybackslash}p{2cm}|}
\hline
\textbf{Representative Words} & \textbf{Human-Readable Label} & \textbf{Number of Conversations and As \% of The Dataset} \\ \hline
['people', 'just', 'dont', 'like', 'think', 'im', 'make', 'youre', 'thats', 'say'] & Residual Topic & 4442 (35.2 \%) \\ \hline
['people', 'dont', 'white', 'trump', 'just', 'black', 'think', 'like', 'im', 'say'] & Trump and Race & 2300 (18.2 \%) \\ \hline
['women', 'gender', 'people', 'men', 'dont', 'trans', 'like', 'sex', 'just', 'think'] & Gender & 1240 (9.8 \%) \\ \hline
['people', 'money', 'government', 'just', 'war', 'dont', 'world', 'think', 'like', 'tax'] & Government, War and Taxes & 929 (7.4 \%) \\ \hline
['child', 'abortion', 'fetus', 'life', 'parents', 'children', 'dont', 'mother', 'people', 'pregnancy'] & Abortion & 608 (4.8 \%) \\ \hline
['rape', 'consent', 'sexual', 'victim', 'sex', 'women', 'raped', 'people', 'assault', 'dont'] & Sexual Violence & 378 (3.0 \%) \\ \hline
['animals', 'meat', 'animal', 'eat', 'dog', 'food', 'eating', 'dogs', 'vegan', 'humans'] & Veganism & 353 (2.8 \%) \\ \hline
['school', 'students', 'college', 'teachers', 'education', 'schools', 'math', 'degree', 'teach', 'job'] & Education & 345 (2.7 \%) \\ \hline
['gun', 'guns', 'police', 'weapons', 'people', 'firearms', 'shootings', 'dont', 'assault', 'weapon'] & Gun Violence & 291 (2.3 \%) \\ \hline
['alcohol', 'drink', 'car', 'drinking', 'drunk', 'drugs', 'driving', 'people', 'cars', 'drug'] & Drugs and Alcohol & 287 (2.3 \%) \\ \hline
['music', 'art', 'movie', 'movies', 'like', 'good', 'film', 'song', 'just', 'characters'] & Art and Movies & 280 (2.2 \%) \\ \hline
['fat', 'weight', 'people', 'depression', 'life', 'suicide', 'just', 'person', 'suffering', 'feel'] & Depression, Suffering, Suicide & 257 (2.0 \%) \\ \hline
['universe', 'life', 'free', 'planets', 'earth', 'probability', 'space', 'quantum', 'brain', 'time'] & The Universe and Space & 144 (1.1 \%) \\ \hline
['game', 'games', 'phone', 'apple', 'play', 'pc', 'pokemon', 'player', 'gaming', 'console'] & Gaming & 131 (1.0 \%) \\ \hline
['piracy', 'content', 'ads', 'money', 'product', 'copyright', 'free', 'music', 'dont', 'pirate'] & Copyright and Piracy & 90 (0.7 \%) \\ \hline
['women', 'sports', 'compete', 'sport', 'men', 'military', 'testosterone', 'combat', 'womens', 'physical'] & Gender, Sports and Military & 82 (0.6 \%) \\ \hline
['coffee', 'pizza', 'cheese', 'cake', 'pie', 'sandwich', 'bread', 'mcdonalds', 'taste', 'cakes'] & Food & 75 (0.6 \%) \\ \hline
['tip', 'tipping', 'service', 'tips', 'pay', 'wage', 'server', 'restaurant', 'servers', 'work'] & Tipping Culture & 57 (0.5 \%) \\ \hline
['bathroom', 'bathrooms', 'toilet', 'unisex', 'shower', 'paper', 'seat', 'pee', 'restrooms', 'urinals'] & Restrooms & 50 (0.4 \%) \\ \hline
['marriage', 'married', 'divorce', 'spouse', 'relationship', 'legal', 'relationships', 'spouses', 'commitment', 'couples'] & Marriage, Divorce and Relationships & 45 (0.4 \%) \\ \hline
['iq', 'language', 'intelligence', 'english', 'languages', 'test', 'learning', 'people', 'mensa', 'smart'] & Intelligence & 38 (0.3 \%) \\ \hline
['israel', 'palestinians', 'jews', 'palestine', 'israeli', 'palestinian', 'state', 'land', 'jewish', 'israelis'] & Israel-Palestine & 38 (0.3 \%) \\ \hline
['football', 'sports', 'team', 'teams', 'trophy', 'brady', 'sport', 'soccer', 'baseball', 'game'] & Sports & 34 (0.3 \%) \\ \hline
['bullying', 'bullied', 'bullies', 'kids', 'bully', 'school', 'schools', 'kid', 'teachers', 'students'] & Bullying & 27 (0.2 \%) \\ \hline
['batman', 'superman', 'hes', 'gotham', 'clark', 'kill', 'responsibility', 'joker', 'world', 'humanity'] & Superheroes & 22 (0.2 \%) \\ \hline
['flight', 'plane', 'airline', 'seat', 'seats', 'flights', 'wright', 'fly', 'aircraft', 'planes'] & Aircrafts & 19 (0.2 \%) \\ \hline
['celsius', 'metric', 'fahrenheit', 'insurance', 'scale', 'temperature', 'water', 'imperial', 'measurements', 'useful'] & Metrics Versus Imperial Systems & 17 (0.1 \%) \\ \hline
['chopsticks', 'fork', 'knife', 'dishes', 'asian', 'rice', 'eating', 'bone', 'eat', 'food'] & Cutlery and Dining & 14 (0.1 \%) \\ \hline
['vim', 'editor', 'tools', 'text', 'code', 'use', 'debugger', 'editors', 'ruby', 'java'] & Coding and Code Editors & 13 (0.1 \%) \\ \hline
['tattoo', 'tattoos', 'meaning', 'people', 'dont', 'body', 'person', 'really', 'think', 'just'] & Tattoos & 13 (0.1 \%) \\ \hline
['analog', 'clocks', 'clock', 'digital', 'hand', 'hour', 'minutes', 'time', 'minute', 'batteries'] & Clocks and Batteries & 11 (0.1 \%) \\ \hline
\textbf{Total} & & \textbf{12,630 Conversations} \\ \hline
\end{tabular}
\caption{Topic Representation and Human-Readable Labels}
\endgroup
\label{table:topic_representation}
\end{table}

Figure~\ref{fig:topic_distribution} shows how constructive and destructive outcomes are distributed amongst the topics. We observe that some topics never resulted in a destructive outcome. These include ``Clocks and Batteries,'' ``Coding and Code Editors,'' and ``Cutlery and Dining.'' However, these topics also represented a very small portion of the dataset, with only 11, 13, and 14 conversations about these topics, respectively.

Conversely, some topics were overwhelmingly likely to lead to a destructive outcome: ``Israel-Palestine'' and ``Race and Trump'' were among the most likely to result in personal attacks. Among the 2,300 conversations about Race and Trump, 67\% of them resulted in a negative outcome --- likely an important reason why this was the most predictive topic label among the 31 categories.

These general distributions align with intuition --- conversations about mundane subjects are relatively unlikely to become heated, while conversations about political issues are more likely to veer off the rails.

Interestingly, however, we find that a constructive and destructive outcome were relatively equally likely for some ostensibly controversial political topics (such as Abortion and Sexual Violence).

\clearpage 

\end{document}